# The exponential distance rule based network model predicts topology and reveals functionally relevant properties of the Drosophila projectome

Balázs Péntek[1,2], Mária Ercsey-Ravasz[1,2,*]

1 Faculty of Physics, Babeş-Bolyai University, Cluj-Napoca, Romania
2 Transylvanian Institute of Neuroscience, Cluj-Napoca, Romania
*Correspondence: maria.ercsey@ubbcluj.ro;

## SUMMARY

Studying structural brain networks has witnessed significant advancement in recent decades. Findings have revealed a geometric principle, the exponential distance rule (EDR) showing that the number of neurons decreases exponentially with the length of their axons. An EDR based network model explained various characteristics of inter-areal cortical networks in macaques, mice, and rats. The complete connectome of the Drosophila fruit fly has recently been mapped at the neuronal level. Our study demonstrates that the EDR holds true in Drosophila, and the EDR model effectively accounts for numerous binary and weighted properties of neuropil networks, also called projectome. Our study illustrates that the EDR model is a suitable null model for analyzing networks of brain regions, as it captures geometric and physical constraints in very different species. The importance of the null model lies in its ability to facilitate the identification of functionally significant features that are not caused by inevitable geometric constraints, as we illustrate with the pronounced asymmetry of connection weights important for functional hierarchy.

## INTRODUCTION

In experimental and theoretical studies, much attention has been paid in the last few decades to understanding the properties of the complex structural network of the brain, because these are key to its functioning. Experiments have been focusing mainly on mapping these structural brain networks in different species. In mammals the mapping is usually performed on meso- or macro-scale: finding the connections between functional areas. Great advances have been made with retrograde tracing experiments in macaque (Markov et al., 2011, 2014) and anterograde and retrograde tracing in mice (Oh et al., 2014; Zingg et al., 2014; Horvát et al., 2016; Gămănuţ et al., 2018). In humans only non-invasive techniques (e.g. DTI) are available, however, the precision of these methods is still investigated (Dyrby et al., 2007; De Reus & Van Den Heuvel, 2013; Thomas et al., 2014; Knösche et al., 2015; Donahue et al., 2016; Maier-Hein et al., 2017). Mapping the brain on neuronal level has started with C. Elegans (White et al., 1986) and recently the whole brain of the Drosophila fruit fly has been mapped (Zheng et al., 2018; Heinrich et al., 2018; Buhmann et al., 2021; Dorkenwald et al., 2023; Schlegel et al., 2023), the database containing a great amount of information about the approx. 140,000 neurons and 34 million synapses included in 78 neuropils (larger brain areas defined in this map)(Dorkenwald et al., 2023; Schlegel et al., 2023).

Conceptualizing interareal neuroanatomy in terms of graphs (Bullmore & Sporns, 2012; Sporns, 2000, 2011) brought the large tool set of graph theory (Barabási & Pósfai, 2016; Newman, 2010) handy for analyzing cortico-cortical graphs in mammals. Interesting features related to wiring optimization have been identified, such as the small-world topology (Bassett & Bullmore, 2017; Watts & Strogatz, 1998), which balances local specialization and global integration. Additionally, hierarchical modularity has been discovered (Bullmore & Sporns, 2012; C. Hilgetag et al., 2000; C. C. Hilgetag & Kaiser, 2004; Meunier et al., 2010; Sporns & Betzel, 2016; Van Den Heuvel et al., 2012) indicating that cortical networks are organized into modules with dense intra-modular connections and sparser inter-modular connections, facilitating efficient information processing and communication. Later, based on the empirical observation of the exponential distance rule in the macaque brain, a predictive network model has been defined (Ercsey-Ravasz et al., 2013), that explained many topological features of these networks, also providing deeper explanations to wiring optimization, modular structure and efficient information processing.

The exponential distance rule states that the probability of axons with a given length decreases exponentially as function of their length: $p(d) = \lambda \cdot \exp(-\lambda \cdot d)$. Simply saying, there are many neurons with short axons and only a few with long axons, this decay being exponential. This was first observed in the white matter of the macaque (Ercsey-Ravasz et al., 2013), later in the mouse (Gămănuţ et al., 2018; Horvát et al., 2016). It was shown to be true also in the gray matter of the macaque and mouse (Horvát et al., 2016). A recent study suggests it also holds in marmosets, humans, and Drosophila (Józsa et al., 2024). It appears to be a basic geometrical principle that achieves the balance between wiring optimization and efficient communication. The $\lambda$ decay parameter is different in species depending on brain sizes.

Recent theoretical studies (Pósfai et al., 2024) also demonstrate that in physical networks where nodes and links are placed in geometrical space, have physical size and cannot intersect, in a densely packed state the exponential distribution of link lengths follows naturally from these geometrical constraints. These theoretical results make us expect the exponential distance rule - as observed before in macaque and mouse - to be true probably in any brain.

A one-parameter predictive network model (Ercsey-Ravasz et al., 2013) was able to explain many important properties of the inter-areal cortical network in macaque and mouse (Gămănuţ et al., 2018; Horvát et al., 2016). In rats the model was tested on two different scales (Noori et al., 2017), appearing to be valid only on a larger scale. However, this data was a large collection based on the bibliography of the last 50 years (Noori et al., 2017), and information is still missing, so the validity of the model on the lower scale remains an open question.

The new experiments that mapped the whole connectome of Drosophila on neuronal level (Buhmann et al., 2021; Dorkenwald et al., 2023; Heinrich et al., 2018; Schlegel et al., 2023; Zheng et al., 2018) raise the questions if 1) the EDR is true in Drosophila brain as expected based on previous empirical results and recent theoretical arguments and what is the decay parameter $\lambda$ in the Drosophila?; 2) Does the one-parameter EDR network model also work in the Drosophila and what properties does it reproduce?

First, we will demonstrate the validity of the exponential distance rule. The FlyWire database allowed us to extract a corresponding graph for approx. 130.000 neurons, on which we are able to measure the cable length between the soma and pre-synapses, the distributions clearly indicating the presence of EDR. Having the coordinates of soma and all synapses we also calculate Euclidean distances, showing how strongly the two correlate with each other and what is the scaling factor between them. This is needed because we are able to construct the EDR model only by using Euclidean distances. This will be done on the level of neuropils (areas). We build the network using the method introduced by Dorkenwald and colleagues (Dorkenwald et al., 2023; Lin et al., 2023). Neuropils also have their coordinates in the JFRC2 Template Brain dataset allowing us to build the Euclidean distance matrix between neuropils (Jenett et al., 2012). Using this we apply the one-parameter EDR network model first introduced by Ercsey-Ravasz et al. (Ercsey-Ravasz et al., 2013) and compare its properties to the real dataset at different parameter values. We show that the optimal parameter reproducing a whole range of network properties coincides well with the parameter measured directly from the exponential distance rule.

The success of the EDR model across species from various evolutionary branches (insects, rodents, primates) suggests that it is the appropriate null model for comparing structural brain networks in future studies. The significance of the null model stems from the need to compare experimental data with models that incorporate unavoidable geometric features present in physical networks, such as the EDR model. This kind of comparisons can more easily reveal functionally relevant features that are not directly attributable to geometric factors. As we will demonstrate, a compelling illustration of these unique properties, not attributable to simple geometric factors, is the pronounced asymmetry in link weights between bidirectionally connected area pairs. This asymmetry plays a crucial role in determining the functional hierarchy of brain areas.

## RESULTS

### The Drosophila database

The fruit fly (Drosophila melanogaster) has been used as a model organism in biology since the early 20th century. From a neuroscience standpoint, it has become an ideal experimental subject due to its rich functionality. In addition to basic functions like vision, flight, and walking, the fruit fly exhibits more complex behaviors such as courtship and aggression (Coen et al., 2014; DasGupta et al., 2014; Dorkenwald et al., 2022; Duistermars et al., 2018; Jennings, 2011; Owald et al., 2015; Seelig & Jayaraman, 2015).

In 2018, a breakthrough occurred in the field of connectomics when Zheng and colleagues successfully mapped the entire brain of the fruit fly (Zheng et al., 2018). Researchers developed a serial section transmission electron microscope (ssTEM) that allowed for the mapping of the entire adult fruit fly brain at the synaptic level, with nanometer resolution (see details in Materials and methods section). In 2021, Julia Buhmann and her colleagues (Buhmann et al., 2021) succeeded in training a convolutional neural network with tens of millions of parameters capable of recognizing pre- and post-synaptic points without the need for reconstructed neurons. Their method allows for determining whether a voxel is a postsynaptic side and, if so, calculating a vector pointing to the presynaptic side. Within the connectome, the authors claim that connections within the network can be determined with approximately 95% accuracy by filtering out connections with fewer than 5 synapses (Buhmann et al., 2021).

The FlyWire project was launched around the beginning of 2022 (Dorkenwald et al., 2022). As described in the Materials and methods section, we downloaded the data from the Codex platform (the version of the database: snapshot 783 - Oct 2023), which contains 139,255 validated neurons, 2,700,513 connections, and 34,153,566 synapses. The database administrators specifically noted that the 2.7 million connections between neurons represent a subset of all detections, and these connections contain more than 4 synapses. This threshold value coincides with the value introduced by Buhmann and colleagues for synapse prediction. In addition to predicting these filtered connections with very high accuracy, the use of the threshold value can also be argued from a biological perspective. Stronger (multi-synaptic) connections may play a more important role in communication between neurons, and it is more likely that these are independent of individual differences (Dorkenwald et al., 2023). We exclusively analyzed intrinsic neurons, totaling around 118,000 (85%), excluding afferent and efferent neurons which may have different structures due to their roles in sending or receiving information outside the brain (Dorkenwald et al., 2023).

## The Exponential Distance Rule in Drosophila

In order to verify the exponential distance rule, we needed to measure the axon lengths. While a recent study (Józsa et al., 2024) has shown the presence of the EDR in Drosophila, they used the cable length values provided in the database, which is defined as the total sum of distances between neighboring nodes in the neuron tree (including the whole tree structure of the axon and dendrite). Here we need to precisely determine the λ decay parameter corresponding to the main axon lengths, therefore, we downloaded spatial graph/tree structures from the Codex web interface (Matsliah et al., 2023), which correspond to the "skeleton" of real neurons (their digital representation). Using these neuron trees, we measured the axon lengths as follows: 1) we identified the node (super voxel) marked as the soma and determined all voxels indicating the presynaptic points of that neuron (synapses between neurons with connections below the 5-synapse threshold are not considered), 2) given one pre-synaptic point we measure the axon length by calculating the shortest path between the soma and the given point on the tree structure of the neuron downloaded from the database, 3) We calculate these distance for all pre-synaptic points of a neuron, we identify the closest pre-synaptic point and we also calculate the average for all pre-synaptic points.

We select the nearest pre-synaptic point to the cell body because we are interested in the main axon projection itself. After the closest pre-synaptic point, the axon almost always branches into many directions (see Figure 1A). It should also be mentioned that for certain neurons, this nearest pre-synaptic point is not necessarily located around the end of the axon, but it may be much closer on the axon, or even on the dendrite. There are two possible reasons for this: the synapse-predicting neural network makes errors, or these points may represent non-traditional synapses, such as dendrite-dendrite or axon-axon connections (Dorkenwald et al., 2023; Eckstein et al., 2024; Galindo et al., 2023; Meinertzhagen, 2018; Schneider-Mizell et al., 2016; Winding et al., 2023) (Supplementary Figure 1A). To eliminate these errors and special cases, we calculate the difference between the path length of the closest pre-synaptic point and the average path length to all pre-synaptic points of that neuron. The distribution of these differences is shown in Supplementary Figure 1B for all neurons. We use a threshold of 0.15 mm to eliminate neurons where errors are most probable and here we will plot the exponential distance rule for the remaining 105,216 neurons (only 11% of neurons are eliminated from the statistics).

Figure 1B,C shows the probability distribution of axonal path lengths for these neurons considering the minimum path length and average path length. The histograms show the probability of an axon having a given length. Setting the $y$ axes on log scale, we can indeed observe the presence of the exponential decay. Changing the bin size may slightly change the fitted parameter of the exponential distribution. We are mainly interested in the decay rate provided by the minimum path length approximated in the $\lambda_{d_{min}} = [15.6, 19.8]\ \mathrm{mm}^{-1}$ interval, but we can see that the average path lengths give similar values ($\lambda_{<d>} = [18.6, 20.2]\ \mathrm{mm}^{-1}$).

The distances between neuropils are Euclidean distances calculated based on their coordinates. For this reason, after calculating axonal path lengths we also calculate Euclidean distances between the soma and presynaptic points (see Figure 1A). As expected, there is a strong correlation between the two, identifying a relatively large scaling factor of approximately 1.58-1.77 (Figure 1D,E). This is an important difference compared to mammals. Because of the small brain size of the fruit fly, the axons can be extremely long compared to the distances between neuropils identified. This difference is not as significant in the larger brains of macaques or even mice, where one can more easily estimate and measure the paths of axon bundles going through the white matter between functional areas. In the fruit fly each neuron has an independently complex tree structure, and axons do not necessarily travel through the brain in a direct, relatively straight path.

In Figure 1E,F we plot the EDR for the Euclidean distances measured, both for the minimum length and the average length. In this case we obtain an interval of $\lambda_{ED_{min}} = [31.1, 35.0]\ \mathrm{mm}^{-1}$ for the minimum length and $\lambda_{<ED>} = [31.6, 37.0]\ \mathrm{mm}^{-1}$ for the average lengths.

The EDR distributions for all neurons without using a threshold for the distance between minimum and average path length are shown in Supplementary Figure 2. We can see that the difference is not large, nevertheless, the EDR is clearer when eliminating the errors.

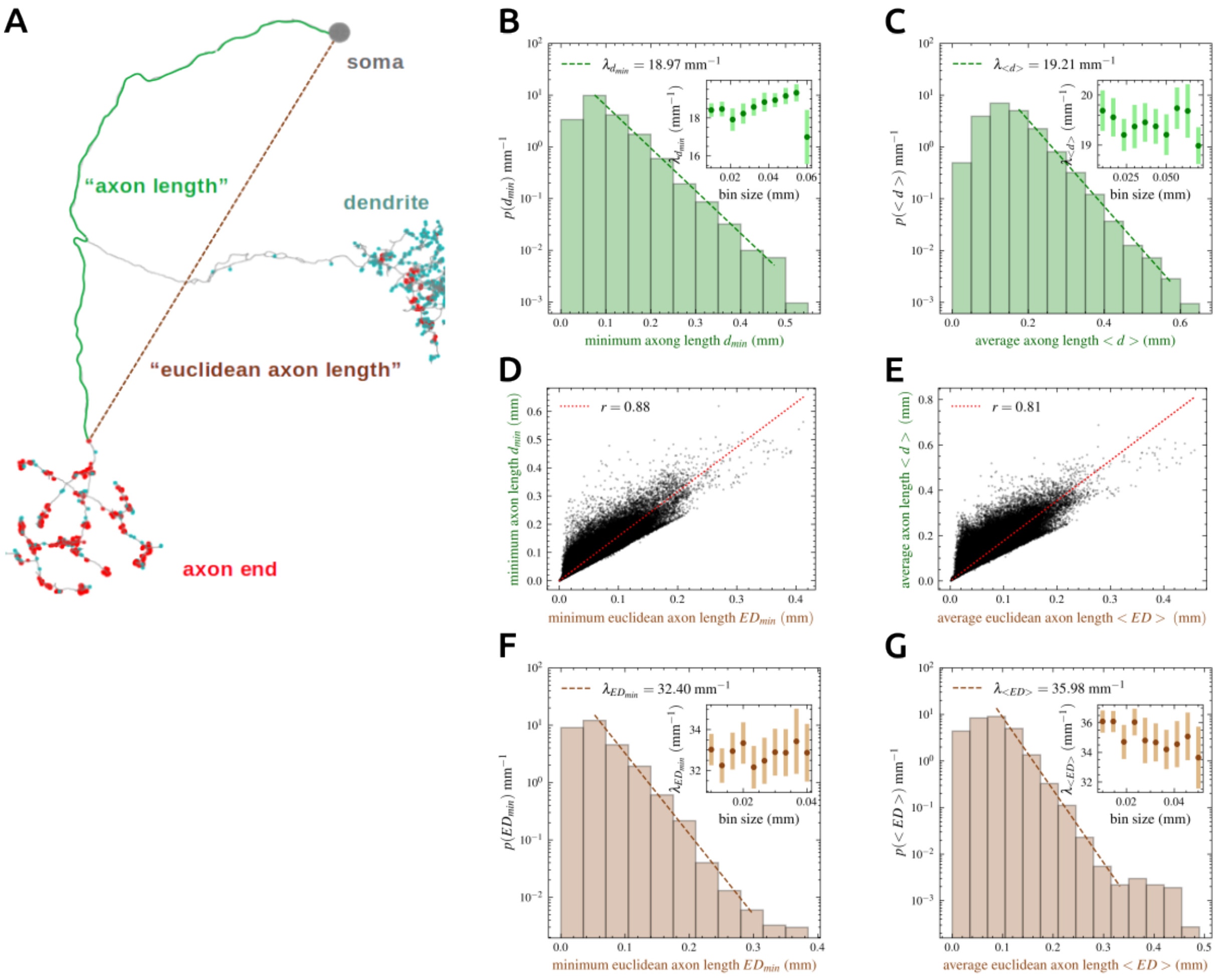

**Figure 1. Measuring axonal length.** A) Tree of a neuron with presynaptic (red) points and postsynaptic (blue) points identified. The green path shows the minimal axonal length, the brown line indicates the minimal Euclidean axon length. B) Probability distribution of minimal axon lengths showing the EDR. Using different bin sizes we estimate $\lambda_{d_{min}} = [15.6, 19.8]\ \mathrm{mm}^{-1}$. C) EDR for average axonal path length provides $\lambda_{<d>} = [18.6, 20.2]\ \mathrm{mm}^{-1}$. D) Correlation between path length and Euclidean distances for minimum and E) average path length values. Scaling factors are 1.58 and 1.77. F) EDR of Euclidean axon lengths for minimum and G) average values providing $\lambda_{ED_{min}} = [31.1, 35.0]\ \mathrm{mm}^{-1}$ and $\lambda_{<ED>} = [31.6, 37.0]\ \mathrm{mm}^{-1}$ respectively.

## The network of neuropils in Drosophila

The brain of the fruit fly is much denser ($6.9\ \mathrm{synapses}/\mu\mathrm{m}^3$) than that of mammals ($< 1\ \mathrm{synapse}/\mu\mathrm{m}^3$), and the structure of most of its neurons differs from what is typical in mammals, in the sense that the soma and dendrites are most of the time spatially separated (Dorkenwald et al., 2023). The soma are mainly located on the surface of the brain, with a primary neurite penetrating into the interior of the brain, where it then branches into two parts: dendrite and axon (for example see Figure 1A). Therefore, in the case of the fruit fly, associating a single neuron with a single brain region is not possible.

Although neurons cannot be associated, it is possible to associate synapses to brain regions. This was successfully achieved by Dorkenwald and colleagues, utilizing a previously determined atlas of neuropils (brain regions) from a prior study (Ito et al., 2014), categorizing synapses into 78 zones based on their pre-synaptic sides. The same team even developed a method of defining the projectome on the level of neuropils (Dorkenwald et al., 2023; Lin et al., 2023). We used the same method, as described shortly in Materials and methods section. The matrix elements obtained actually represent a strength value, characterizing the connections between neuropils (brain regions). Considering that brain regions can vary in size (see Figure 2A), the numbers of synapses can differ significantly between larger and smaller neuropils. Therefore, we normalized the obtained connectivity matrix by columns, so that the newly obtained $w_{ij}$ matrix element will correspond to the probability of information from area $i$ flowing to area $j$. This way it has similar meaning to the fraction of labeled neurons (FLN values) applied in retrograde tracing studies in macaque and mice (Ercsey-Ravasz et al., 2013; Horvát et al., 2016), where also neuron counts are normalized separately for each injection.

To calculate the distances between brain regions, we used the locations of mass centers published in an earlier study (Jenett et al., 2012), obtained through the fafbseg.py Python package. These coordinates were available for a total of 75 brain regions, as the FlyWire team added three brain regions to the database later on. For our purposes, this did not represent a significant loss, as these three zones are already close to sensory organs, and we are mainly interested in intra-brain connections (analyzing only intrinsic neurons), as mentioned above. Two of these zones are associated with the left and right eyes (lamina of the compound eyes - see the two outer magenta zones in Figure 2A), while the third is the ocellar ganglion located on top of the head (see the small pink sphere at the top in Figure 2A) playing a role in flight and spatial orientation. Therefore, we recalculated the projectome only for these $N = 75$ brain regions (neuropils), resulting in a network with $M = 4733$ connections (density: 85%).

In Figure 2B we plot the distribution of Euclidean distances between neuropils, which can be well approximated with a truncated Gaussian distribution. This is somehow expected from the spatial positioning of neuropils. The distribution of normalized connection weights ($w_{ij}$) shows a lognormal distribution ranging over several orders of magnitudes. As first explained in the case of the macaque, the Gaussian distribution of distances combined with the exponential distance rule gives a theoretical support for the lognormal distribution of connection weights (Ercsey-Ravasz et al., 2013; Horvát et al., 2016). As we see, this is not different in case of the fruit fly.

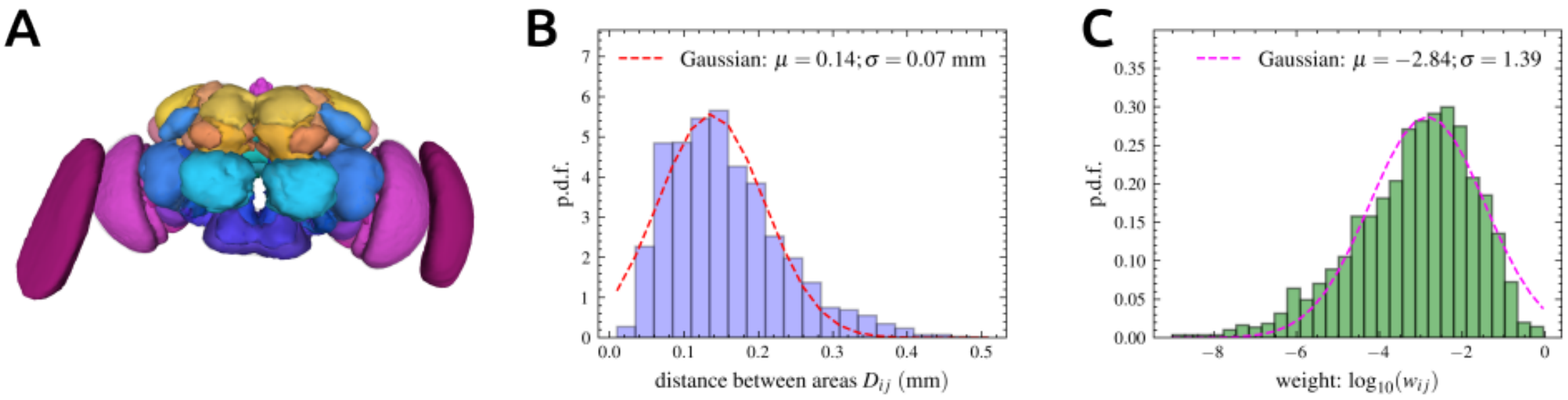

**Figure 2. The neuropil network**. A) The atlas of neuropils, image obtained from the *Connectome Data Explorer* (*Codex*) web-app (codex.flywire.ai/app/neuropils). B) Distribution of Euclidean distances between neuropils. C) Distribution of logarithmic weight values ($\log_{10} w_{ij}$ ) is a normal distribution, indicating that the weight distribution is lognormal.

We investigated the modularity of the network with a hierarchical clustering method (Murtagh & Legendre, 2014; Ward, 1963) (see Materials and methods section), the weighted connectivity matrix with the dendrogram and the 4 largest modules colored are shown in Figure

3. The clustering provides a realistic modular structure with symmetric organization and clusters localized in space.

The first big cluster (blue, Figure 3B,C) contains mainly the left and right optic lobe with the Medulla (ME), Accessory Medulla (AME), Lobula (LO) and Lobula Plate (LOP) and interestingly also includes the antennal mechanosensory and motor center (AMMC) from the periesophageal regions.

The second cluster (orange, Fig. 3B,C) is located mainly at the inferior and back side of the brain. It includes the Lateral Accessory Lobe (LAL); the Inferior Clamp (ICL) and Inferior Bridge (IB) from the inferior protocerebrum; all ventromedial regions: Vest (VES), Superior Posterior Slope (SPS), Inferior Posterior Slope (IPS), Epaulette (EPA), and Gorget (GOR); almost the whole ventrolateral part: Anterior and Posterior Ventrolateral Protocerebrum (AVLP, PVLP), Posteriolateral Protocerebrum (PLP), and Wedge (WED); the Gnathal Ganglia (GNG) and interestingly two right side periesophageal regions: right Flange (FLA_R) and Cantle (CAN_R).

The third cluster (green, fig. 3B,C) contains a large part of the central complex: the Ellipsoid body (EB) and Noduli (NO), and surprisingly even if they are not neighbors in space it also includes the left and right Bulb (BU) and Gall (GA) from the lateral complex, meaning there are strong connections between them.

The fourth cluster (red, fig. 3B,C) is localized more at the top and frontal side of the brain. It contains the whole superior protocerebrum: Medial (SMP), Intermediate (SIP) and Lateral (SLP); the rest of the inferior protocerebrum: Antler (ATL), Crepine (CRE) and Superior Clamp (SCL); one single region from the ventrolateral part: the Anterior Optic Tubercle (AOTU); the whole Mushroom Body with the Vertical and Medial Lobe (MB_VL, MB_ML), Pedunculus (MB_PED), and Calyx (MB_CA); the Antenna Lobe (AL); the rest of the central complex: the Fan Body (FB) and Protocerebral Bridge (PB); the Lateral Horn (LH); from the periesophageal regions the Prow (PRW) and only the left Cantle (CAN_L) and left Flange (FLA_L). These two regions (CAN, FLA) are the only ones where the left and right parts are not in the same cluster.

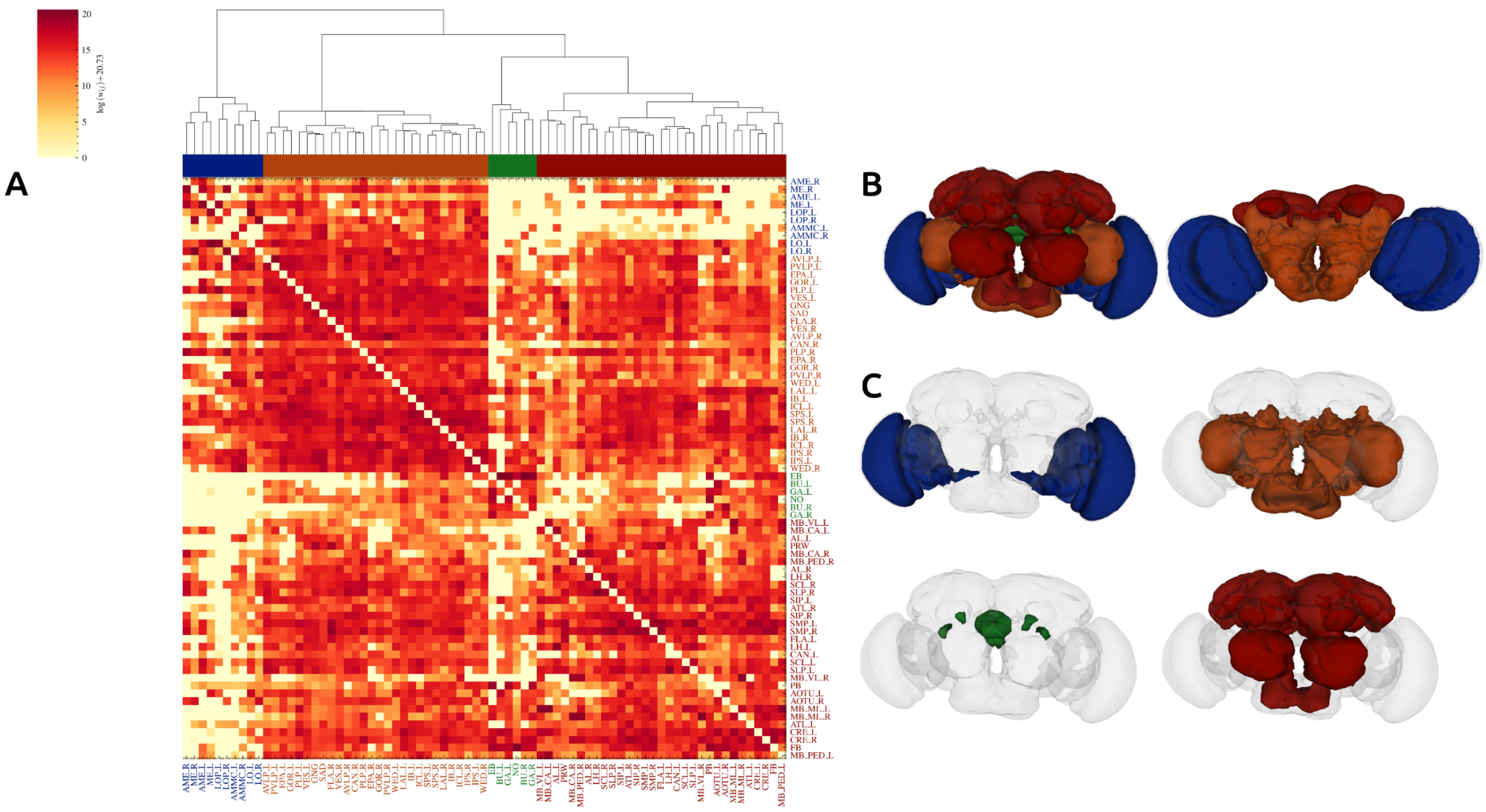

**Figure 3. Modular structure of the Drosophila neuropil network**. A) The weighted connectivity matrix with the areas ordered according to the dendrogram provided by the hierarchical clustering based on Ward's method (see Materials and methods section). We color the four largest clusters under the dendrogram. B) The four largest clusters presented together on the *Drosophila* brain from a front and a back view. C) The four clusters are presented separately from the front view. The blue cluster contains mainly areas from the optic lobe. The orange cluster contains all ventromedial regions, almost the whole ventrolateral part, some parts of the lateral complex and of the inferior protocerebrum, the Gnathal Ganglia and two right side periesophageal regions (FLA_R, CAN_R). The green cluster is concentrated around the central complex (EB and NO) including also parts of the lateral complex. The red cluster includes the whole superior protocerebrum, the mushroom body, the antenna lobe, the lateral horn, some regions from the inferior protocerebrum, ventrolateral part, central complex and periesophageal areas.

**The EDR based network model**

The EDR based network model is a one-parameter maximum entropy based model (Jaynes, 1957) first introduced by Ercsey-Ravasz et al. (Ercsey-Ravasz et al., 2013). The goal is that knowing the distance matrix between brain areas $D_{ij}$ and the number of connections $M$ in the structural network, the model should generate a random network with the same density, that takes into account only the presence of the exponential distance rule with a given decay parameter $\lambda$, but everything else is taken as random. Comparing these random networks with the real data can provide information about the properties of the network that are explained by this geometrical rule.

In our case the input is the distance matrix between the 75 neuropils (obtained from JRFC2 Template Brain dataset (Jenett et al., 2012)), $D_{ij}$ and the number of projections in this network $\mathrm{M} = 4733$. The steps of generating the random network are the following: 1) We randomly choose a distance value based on the exponential distribution with a given parameter $\lambda$; 2) We choose uniformly at random an area pair with distance from the distance bin in the histogram of distances corresponding to the distance chosen in point 1); 3) We choose uniformly at random a direction between the areas and we insert a connection. Multiple connections between areas are allowed generating the connection weights between areas. 4) We stop when the number of binary links between areas reaches $M$. This way we obtain a model network with the same density as the dataset. 5) At the end we normalize the columns of the weighted matrix (the sum of weights in each column will be 1), the weights meaning probabilities of information transfer.

For every $\lambda = 0, 5, 10 \dots 60\ mm^{-1}$ we repeated this procedure generating 1000 random graphs and we measured all types of binary and weighted network properties, comparing them to the properties of the real neuropil network. The definitions of network properties calculated can be found in Materials and methods section. It is important to mention that the 1000 model networks are all analyzed separately, the network properties being calculated for each of them individually. Only later may we apply statistics, such as calculating averages or standard deviation values. One should not consider averaging the weighted networks and building one averaged network, as this may provide a network with a different density and different properties, inducing misleading results (Varga et al., 2024).

## Comparing binary network properties

First, we looked at the degree distributions of nodes. Having a small network with 75 nodes, these are relatively noisy (see Supplementary Figure 3) so here we rather plot the in- and out-degree of nodes in decreasing order (Fig. 3A,B) and compare these with curves generated by the EDR model with $\lambda = 0\ mm^{-1}$ (also called constant distance rule, CDR) and $\lambda = 33\ mm^{-1}$ (EDR), this value being based on the fit in Figure 1F. Figure 3C shows the root mean square deviation (see Materials and methods section) between these curves provided by the data and model as function of the lambda parameter, indicating that the minimum differences are indeed in/very close to the ones in the $\lambda_{ED_{min}} = [31.1, 35.0]\ \mathrm{mm}^{-1}$ (gray interval) provided by the fitting in Fig.1F.

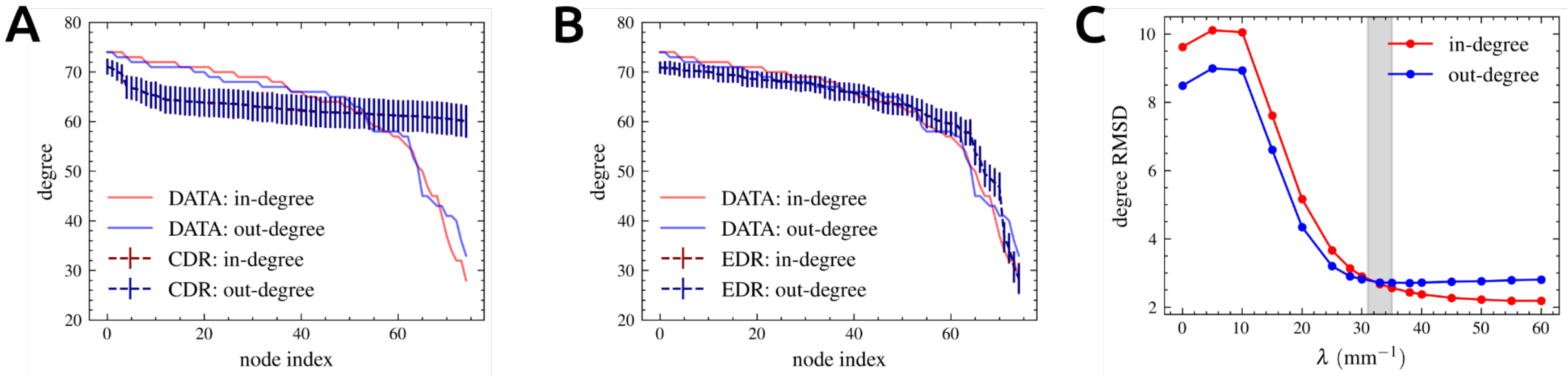

**Figure 4. Node degrees in descending order: dataset vs. model.** A) in- (red) and out-degree (blue) of nodes are shown for each node in decreasing order for the dataset (continuous line), CDR model with $\lambda = 0\ mm^{-1}$, and B) EDR model with $\lambda = 33\ mm^{-1}$. For the models dots represent the average obtained over the 1000 generated networks, vertical lines show the standard deviation. C) RMSD as function of $\lambda$, the gray interval showing the fitted values in Fig. 1F: $\lambda_{ED_{min}} = [31.1, 35.0]\ \mathrm{mm}^{-1}$.

Next, we compared the number of uni- and bidirectional links in the dataset and the model networks generated with different $\lambda$ parameters. In Fig. 5A we can see how the values provided by the model agree with the data exactly in the gray interval based on the fitted $\lambda_{ED_{min}}$ values. Similar observations can be made for the clustering coefficient (Fig. 5B), average binary path length (Fig. 5C) and triangular motifs (Fig. 5D,E,F) (definitions can be found in Materials and methods section). For the triangular motifs we first show the distribution of the 16 possible directed triangular motifs in Fig. 5D, then the relative differences between the data and the average provided by the network models is shown on log scale (Fig. 5E). The root mean square deviation (RMSD) between the data and the model (average) distributions is shown in Fig. 5F, again providing minimal values in the lambda interval (gray) measured in Fig. 1F. Because of the high network density it is expected that the last two motifs are most frequent, but the difference between them is surprisingly large and this is reproduced only by the EDR model, the CDR provides much closer probabilities for the two.

We also counted the number of fully connected subgraphs (cliques) with a certain size. This being a computationally costly procedure we covered only the clique sizes from 3 to 8, and we also searched for the largest ones above 40 going up to 43. In the neuropil network the largest cliques have 43 nodes, there are 31 number of these including in total 53 of brain areas. Together these form an extremely dense core (density 98%), being similar to the core-periphery structure noticed in the macaque (Ercsey-Ravasz et al., 2013; Markov et al., 2013). Comparing this to the random model networks generated with $\lambda = 0\ mm^{-1}$ (CDR) and $33\ mm^{-1}$ (EDR), we see that the CDR model drastically underestimates the number of cliques, while the EDR model gives good estimation in case of small cliques. Nevertheless, for the largest ones even the EDR model cannot reproduce the huge number of cliques. This shows these are specific structures that even if their presence is supported by this geometrical rule, there must be other reasons for which these are so frequent.

**Comparing weighted network properties**

As one can see, most binary network properties are reproduced surprisingly well by this one-parameter EDR network model. The next step is to consider weighted properties. Here, we must take into consideration that connection weights between neurons (and therefore weights between neuropils derived from these values) are not as precise, as they are based on predictions provided by convolutional neural network models. Additionally, the 5-synapse threshold considered may eliminate some true connections, making link weights between neuropils slightly weaker. Indeed, looking at the link weight distribution (specifically the distribution of $\log_{10} w_{ij}$ values in Fig. 6A), we can see that the distribution in the data has a longer tail, including weaker connections (smaller $\log_{10} w_{ij}$ values) than the one reproduced by the $\lambda = 33\ mm^{-1}$ EDR model. Nevertheless, the two distributions are not far apart; their characteristic shape is a normal distribution ($w_{ij}$ being lognormal), and as shown in Fig. 6B, the optimal $\lambda$ predicted would be slightly larger, around 40 $mm^{-1}$.

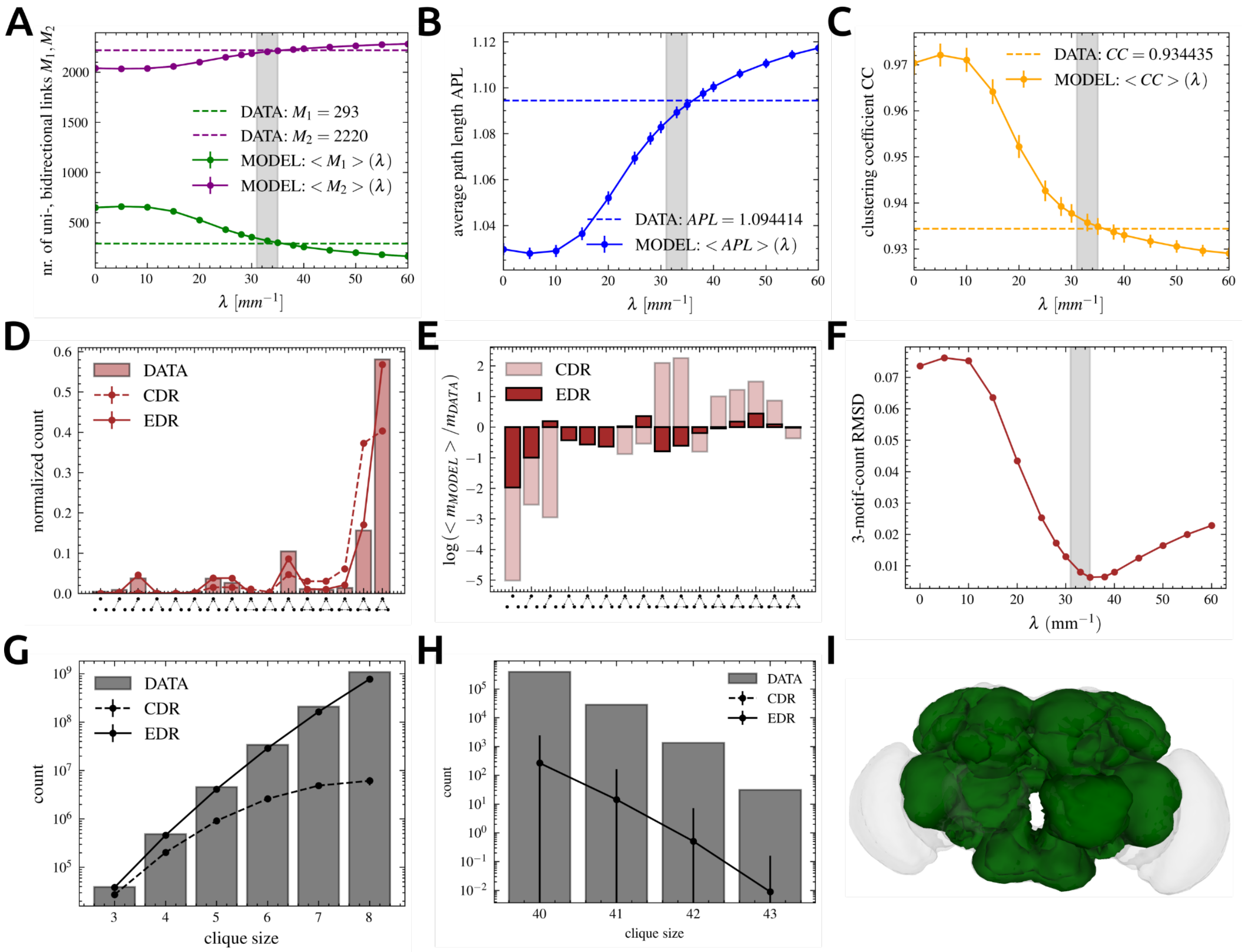

**Figure 5. Comparing the binary properties of the neuropil network with the EDR model.** A) Number of uni- (green) and bidirectional (purple) links as function of $\lambda$. B) Clustering coefficient as function of $\lambda$. C) Average binary path length as function of $\lambda$. Dashed lines show values in the neuropil network. D) Distribution of triangular motifs for data (bars), CDR (dashed line) and EDR with $\lambda = 33\ mm^{-1}$ (continuous line). E) Relative differences between the average obtained from the 1000 model networks and the dataset is shown on log scale. F) RMSD for 3 motif counts as a function of $\lambda$. G,H) Clique size distributions shown for 3 to 8 and 40 to 43, this being the largest clique in the data. I) The total set of 53 areas included in the largest cliques are shown on the brain map (see also Supplementary Figure 4). For the model dots represent the average obtained over the 1000 generated networks, vertical lines show the standard deviation.

Fig. 6 C,D provides a similar illustration for the out-strength distribution of areas. Again, an unnaturally large peak of small strength values is observed in the data. The EDR model reproduces the mean value of the distribution fairly well but provides a shape closer to a normal distribution. However, as shown in Fig. 6 E,F the distribution of distances between nodes (the shortest path length calculated using the $-\log w_{ij}$ length values; see Materials and methods section) is reproduced surprisingly well by the EDR model. We argue that this is because strong connections are well predicted by the algorithms and are more precise. Errors typically have more drastic effects on weaker connections, which do not become part of the shortest paths in the network.

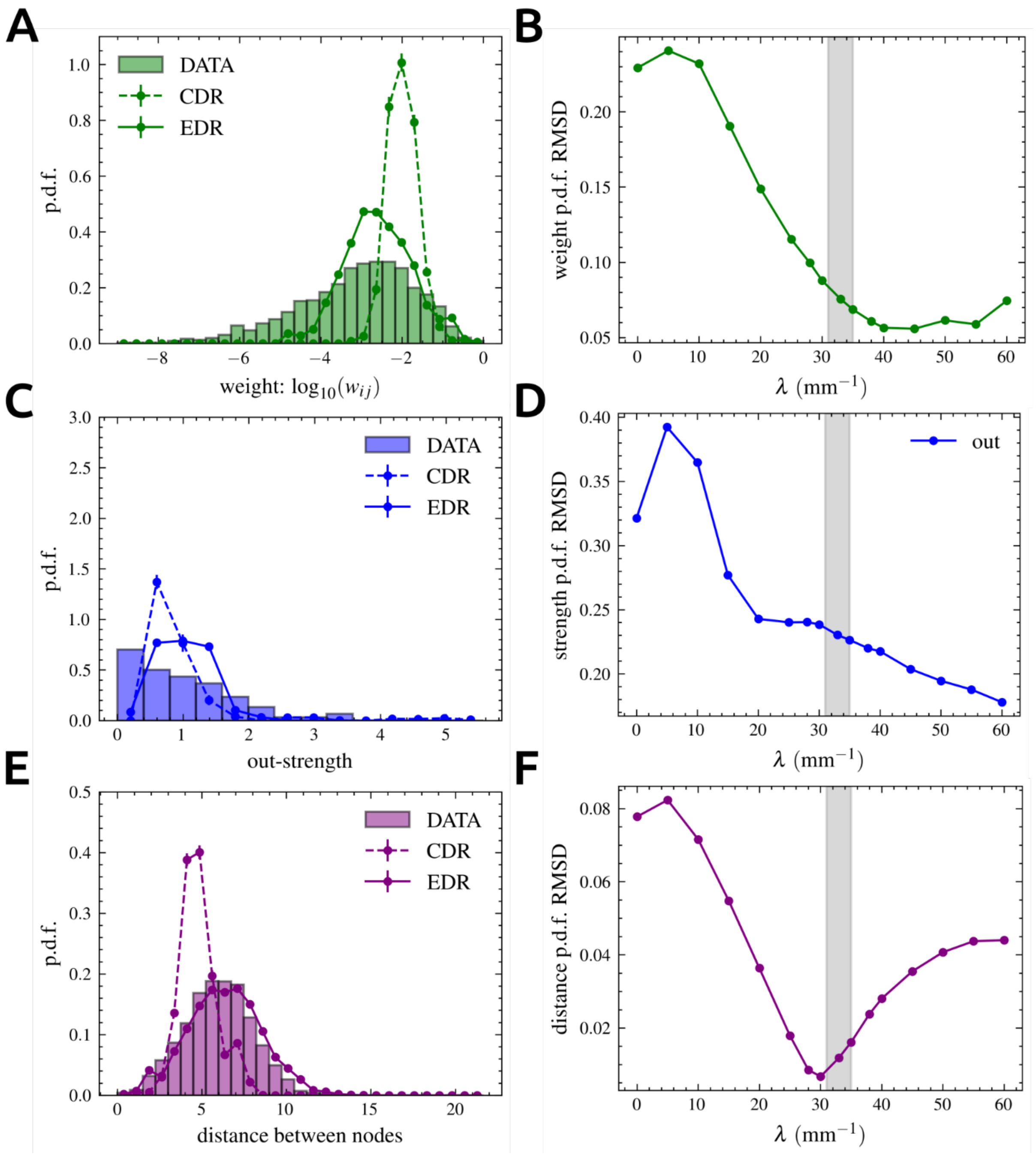

**Figure 6. Comparing weighted network properties of the neuropil network and EDR model.** A) Distribution of logarithmic weight values ($\log_{10} w_{ij}$ ) is slightly wider in the dataset, but clearly the CDR gives a much worse estimation than the EDR model with $\lambda = 33\ mm^{-1}$. B) The RMSD calculated for the link length distributions between data and model is shown as function of $\lambda$. C) Distribution of out-strength (weighted out-degree, see Materials and methods section) of nodes. In-strength is not shown, because it is 1 for each node, the weighted matrix being normalized. D) RMSD for out-strength distributions as function of $\lambda$. E) Distances between all node pairs are calculated in the network using the $l_{ij} = -\log w_{ij}$ length values. Their distribution agrees well with the $\lambda = 33\ mm^{-1}$ EDR model. F) The RMSD calculated between data and model for the node distance distribution as function of $\lambda$.

There are two other network measures based on the shortest path lengths between nodes that characterize the communication efficiency in a network. These are called global and local communication efficiency (Latora & Marchiori, 2003; Vragović et al., 2005) (definitions in Materials and methods section). Similar to the analysis performed by Ercsey-Ravasz et al. (Ercsey-Ravasz et al., 2013) for macaques and Horvat et al. (Horvát et al., 2016) for mice, we plot these efficiency values as a function of network density, while removing links ordered by their weight. In the Drosophila neuropil network, we observe the same behavior as in macaques and mice: when strong connections are removed first, both efficiency values decrease rapidly; when weak connections are removed first, the global efficiency remains almost constant until a low density is reached and the network falls apart, while the local efficiency even increases, showing a large peak at small densities. Similar to macaques and mice, even if the curve is not reproduced precisely, this type of behavior is observed only when applying the EDR model. The CDR ($\lambda = 0\ mm^{-1}$) does not show this phenomenon (Fig. 7A,B). As explained in case of macaque and mice this behavior is supported by the hierarchical modular structure of the network produced by the EDR.

In Figure 7 C,D, we plot the strongly connected backbone of the network (where there are paths in both directions between any pair of nodes) with a density of 35%, and the weakly connected backbone of the network (where there are paths between any pair of nodes, but the directions of links are neglected), with a density of 3%. The figures were produced using the NetworkX software and the Kamada-Kawai algorithm, which is a spring-based force-directed algorithm (Kamada & Kawai, 1989). The nodes have been colored based on their classification into lobes in the FlyWire database. We can see how nodes with the same color usually group together, indicating that structural clusters have functional role in the brain.

**Some functionally relevant properties cannot be reproduced by the geometric model**

Naturally, we cannot expect every property of the projectome to be replicated by a random network model. The importance of the applied null model lies in its ability to facilitate the detection of interesting, functionally relevant properties that are not a direct consequence of physical structure and geometry.

One such property is the asymmetry of link weights between node pairs. For nodes $i$ and $j$, we define this measure as the relative difference of weights in the two directions: $ASYM_{ij} = \frac{|w_{ij}-w_{ji}|}{w_{ij}+w_{ji}}$ (see Materials and methods section). We calculate this measure for all node pairs, but obviously for pairs connected with a link only in one direction, this value is 1. The number of these unidirectional links has been well predicted by the EDR model (Figure 5A). In Figure 8A we show the histogram of the asymmetries only for the bidirectional links. As we can see, there is a large peak at high values (close to 1), and there are few symmetric links with values close to zero. Neither the CDR nor the EDR with $\lambda = 33\ mm^{-1}$ reproduces the characteristic shape seen in the data. These geometric models, being based on the symmetric Euclidean distance matrix between neuropils, do not favor the formation of strongly asymmetric link pairs. The model predicts much lower probabilities for high asymmetry values. Nevertheless, these asymmetries are present in the brain and are probably important in determining the functional hierarchy of brain areas (Felleman & Van Essen, 1991; C.-C. Hilgetag et al., 1996; Hochstein & Ahissar, 2002).

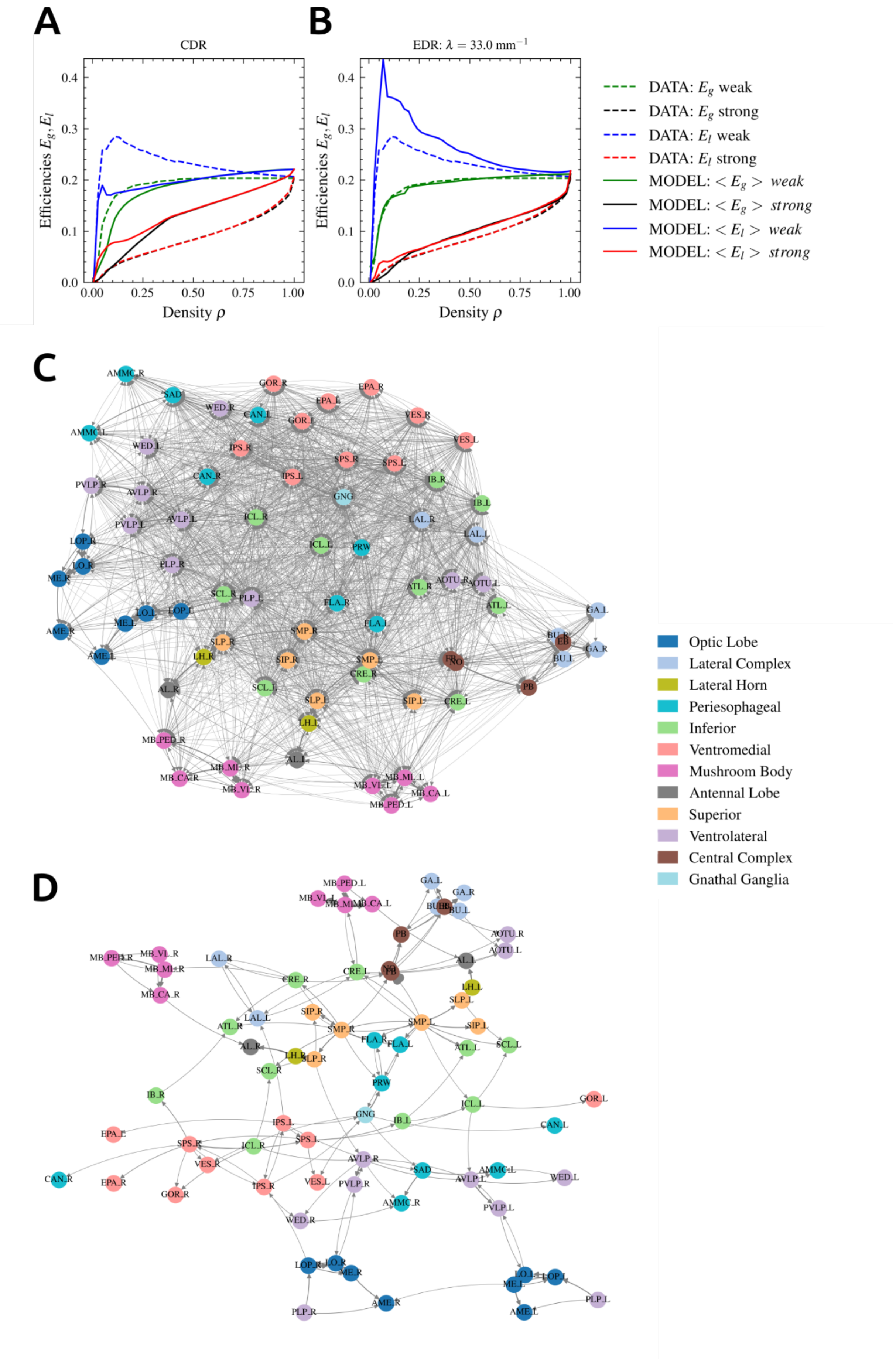

**Figure 7. Communication efficiency and backbone of the projectome**. The global and local communication efficiency values are calculated in the network as function of density when taking out the links one by one ordered by their strength. Two cases are considered, taking out the weakest or the strongest links first (see legends). The results obtained from the data is represented with dashed line, the averages obtained from the 1000 model networks in the A) CDR and B) EDR with $\lambda = 33\ mm^{-1}$ are shown with continuous line. C) The strongly connected and D) weakly connected backbone of the projectome is plotted using the Kamada-Kawai force-layout algorithm.

Fig. 8A is a histogram including all bidirectional links, but we were curious if anything changes when looking at area pairs from the same hemisphere (ipsilateral links, Supplementary Figure 5A) or area pairs from different hemispheres (contralateral connections, Supplementary Figure 5B). These show similar behavior; however, surprisingly, when looking at homotopic connections—link pairs connecting the left and right parts of the same functional areas—the distribution is completely different. These homotopic connections are much more symmetric and are relatively well predicted by the EDR model (Fig. 8B). This supports the idea that this measure can be connected to the functional hierarchy of areas, the left-right parts of the same area being on the same hierarchical level, these are expected to be more symmetric.

In order to illustrate this functional hierarchy, we build a new network characterizing information flow with one single link between each pair of nodes. If $w_{ij} > w_{ji}$ then the link is directed from $i$ to $j$, otherwise from $j$ to $i$, and we use the asymmetry values as link weights of this new network. Calculating the in- and out-strength of areas in this network we characterize how strong is the outgoing/incoming asymmetry. In Figure 8B we color the areas according to their out-strength of asymmetries, showing the map from front, back, top and bottom. There are several observations supporting that this measure gives information about hierarchy: 1) the ranking of areas is strongly symmetrical. 2) We can observe how the areas from the optic lobe (ME, LO) have the largest out-strength being at the lowest level of hierarchy, where information mainly comes in from sensory input and is forwarded to higher level areas for processing (see Supplementary Figure 6A for the order of areas). The areas at the top of hierarchy with lowest out-strength, seem to be CAN, AME, WED, AMMC, MB_VL, etc. Usually areas that are at the bottom of the top list for the asymmetry out-strength are at the top of the list for the asymmetry in-strength (shown on Supplementary Figure 6B), this also supports that this list could give information about the list of hierarchy between areas. The same brain map based on the in-strength of asymmetries is shown in Supplementary Figure 7.

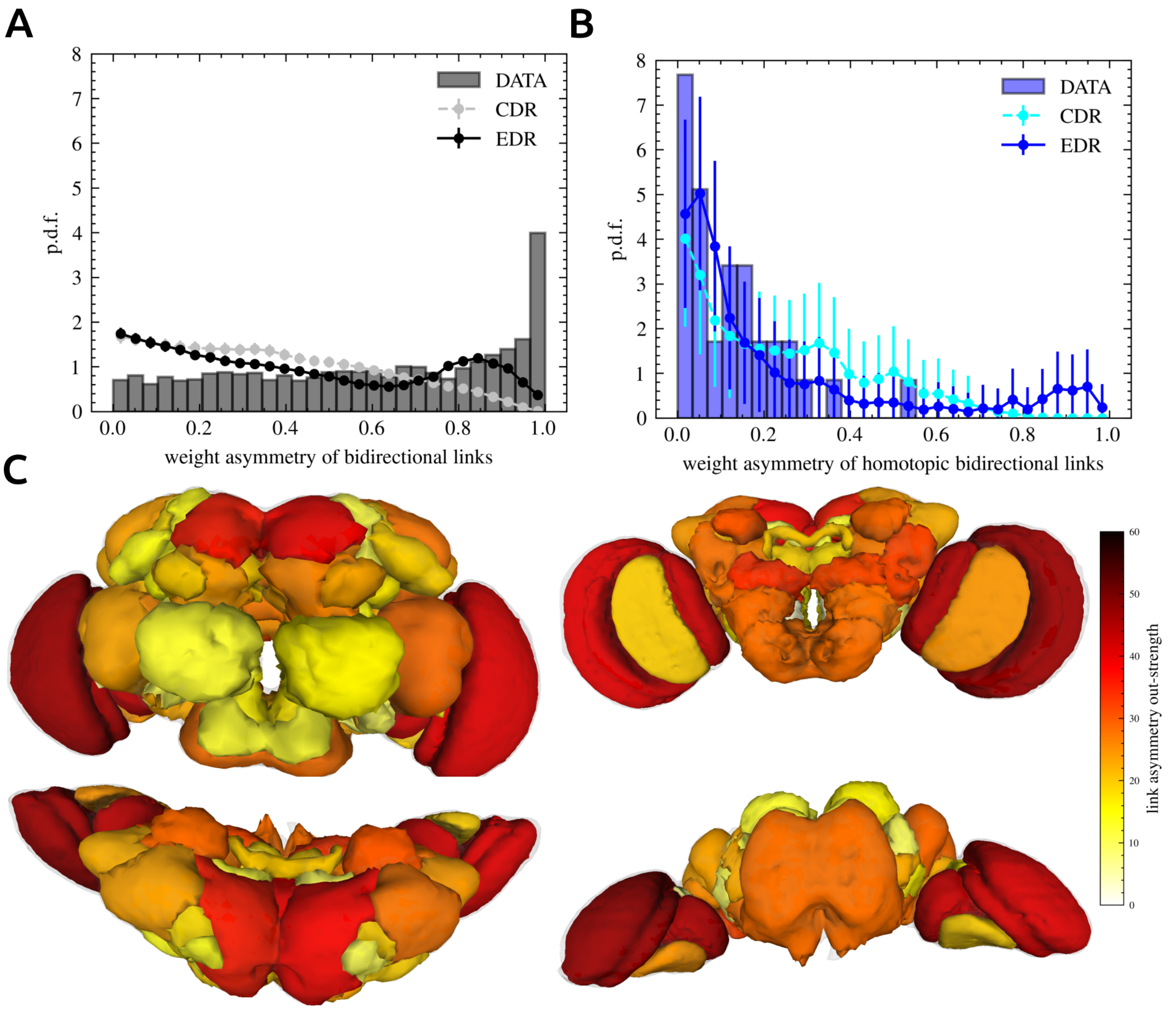

**Figure 8. Asymmetry of connection weights**. A) Probability distribution of asymmetries for all node pairs connected with bidirectional links. B) Probability distribution of asymmetries for homotopic links connecting the same functional regions between left and right hemispheres. C) Brain regions are colored based on the asymmetry out-strength (for definition see Materials and methods section). Front, back, top and bottom view is shown. Areas with high/low asymmetry out-strength are expected to be at the bottom/top of the hierarchy. Similar figure for the asymmetry in-strength is shown in Supplementary Figure 7.

## DISCUSSION

The FlyWire project provides a comprehensive open-access dataset with neuron-level mapping and detailed information on the Drosophila brain. In this paper, we used this data to demonstrate that the exponential distance rule, known to apply in macaques, mice, and rats, also holds true for Drosophila. By analyzing the available neuron tree structures, we measured the decay rate of the exponential distance rule using both real axonal lengths and Euclidean distances. We estimated the decay rate to be in the interval $\lambda_{d_{min}} = [15.6, 19.8]\ \mathrm{mm}^{-1}$ for real axonal paths and $\lambda_{ED_{min}} = [31.1, 35.0]\ \mathrm{mm}^{-1}$ for Euclidean distances. This factor is crucial for understanding Drosophila, whereas in larger brains (such as those of macaques and mice), the difference between these measurements is negligible.

Next, we studied the network of neuropils (brain regions) and applied the EDR-based network model, similar to previous studies on macaques and mice (Ercsey-Ravasz et al., 2013; Horvát et al., 2016). We found that the EDR model accurately predicts most binary properties, such as degree distributions, uni- and bi-directional links, clustering coefficient, average path length, and triangular motifs. However, the model underestimates the huge number of large cliques (completely connected subgraphs). In the neuropil network, the largest clique consists of 43 out of 75 nodes, and there are 31 such cliques, involving 53 nodes in total. These cliques are crucial for the brain's modular structure and hierarchy. Like in macaques and mice, these form a dense core of the network. While the geometric rule supports this property, as the CDR model cannot reproduce even small cliques, the EDR model with $\lambda = 33\ mm^{-1}$ reproduces small cliques but still underestimates the largest ones.

For weighted properties, we delved deeper than previous studies, comparing weight distributions, out-strength and node distance distributions, as well as local and global communication efficiencies. The qualitative behavior is well reproduced, and the behavior of communication efficiency as function of density is similar to that found in macaques and mice. However, there are several quantitative differences in the link weight and out-strength distributions. We suggest this may be due to imprecisions in connection weights between neurons determined by convolutional neural networks and the applied 5-neuron threshold (Buhmann et al., 2021), which can eliminate some true links, weakening connections between neuropils. This argument is supported by the fact that the node distance distribution is accurately reproduced by the model, as shortest paths typically involve strong links that are less affected by errors.

An interesting property of the neuropil network, not directly studied before and not reproduced by the geometrical model, is the asymmetry of weights. By examining all bidirectional links, we calculated the relative difference between the weights in the two directions. While a geometrical model typically does not produce strong asymmetries (being based on the symmetric Euclidean distance matrix between neuropils), these are surprisingly prevalent in the brain and likely important for its functional hierarchy. Another surprising finding is that homotopic connections — links between the left and right sides of the same functional areas — are much more symmetric and align well with the model. This supports the idea that strong asymmetries relate to functional hierarchies, as the left and right sides of the same area are expected to be on the same level of the hierarchy. These asymmetries could be used to develop functional hierarchical models, as attempted before for the visual processing system (Felleman & Van Essen, 1991; C.-C. Hilgetag et al., 1996; Hochstein & Ahissar, 2002). Here, we made a first attempt to build a hierarchy by using the total asymmetry out- and in-strength values to highlight the ranking of areas. This deserves more detailed future studies, but we believe this shows that comparison of real region-level structural brain networks with the EDR-based null model can reveal information about the network's functionally relevant properties. This ability to dissect the nuances of brain structure and function not only enhances our understanding of neural architecture but also opens avenues for future research aimed at exploring the evolutionary implications of these organizational principles across different species.

Previous empirical studies, along with our new research on Drosophila, demonstrate that the exponential distance rule is valid across various species, including insects, rodents, and primates. Theoretical results also support the presence of EDR in densely packed physical networks, such as the brain (Pósfai et al., 2024). Therefore, we argue that the EDR-based model is an appropriate null model for analyzing structural brain networks on meso- and macro-scale (level of brain regions). This model effectively predicts many topological properties, suggesting that most are consequences of geometry and physical structure. Comparing structural networks to the EDR-based null model helps identify functionally relevant features that are not just due to geometry. For example, we highlighted the asymmetry of connection weights, which are likely crucial for forming the hierarchy of functional brain regions. To facilitate future studies, we have made the codes generating EDR model networks available on GitHub (https://github.com/bpentek/EDRmodel).

**ACKNOWLEDGEMENTS**

This study was supported by the Romanian National Authority for Scientific Research and Innovation, CNCS-UEFISCDI, projects: PN-III-P4-PCE-2021-0408 (M. E-R., B.P.), ERANET-FLAG-ERA-ModelDXConsciousness (M.E-R), ERANET-NEURON-2-UnscrAMBLY (M.E-R), ERANET-NEURON-2-IBRAA (M.E-R), ERANET-NEURON-2-RESIST-D (M. E-R., B.P), FLAGERA-JTC2023-MONAD (M E-R, B.P). B.P. was also funded by a student research scholarship of the Babes-Bolyai University nr. 36265/24.11.2023, the Federation of Hungarian Universities from Cluj-Napoca (KMEI) and the Bethlen Gábor Association, and The Bolyai Society from Cluj-Napoca.

**AUTHOR CONTRIBUTIONS**

M.E-R. proposed the research, B.P. analyzed the data and made the figures, M.E-R. supervised the research, M.E-R., B.P. interpreted the results, M.E-R., B.P. wrote the manuscript.

**DECLARATION OF INTERESTS**

The authors have no competing interests.

# MATERIALS AND METHODS

### The Drosophila database

In recent years, the field of connectomics has seen advancements due to improvements in neuroimaging technologies and developments in machine learning/artificial intelligence/neural networks. In 2018, Zheng et al. successfully mapped the connectome of *Drosophila melanogaster* (commonly known as the fruit fly), a model organism in biology since the early 20th century. They developed a specialized serial section transmission electron microscope (ssTEM) that captured images of the adult female *Drosophila* brain with a resolution of just a few nanometers. Their research produced a dataset of approximately 7,000 images, totaling about 106 terabytes, which has been made publicly available (Zheng et al., 2018).

Buhmann et al. utilized this database to predict the chemical synapses in the adult *Drosophila* brain using a large convolutional neural network. The authors found that their model performed best for one-to-many synapses, where a single presynaptic site connects to multiple postsynaptic sites. Considering the polyadic nature of insect synapses (where multiple synapses can connect the same pair of neurons) they determined that setting a threshold of at least 5 synapses/connection allowed for highly accurate predictions of the remaining neural connections (Buhmann et al., 2021). This thresholding can be reasoned also from a more biological perspective, as the stronger connections (with >4 synapses) can be considered physiologically more significant, also present across individuals. (Dorkenwald et al., 2023)

The FlyWire project established by Dorkenwald et al. enabled citizen scientists from all around the world to contribute to the reconstruction of the *Drosophila* connectome, by proofreading automatically traced neurons (Dorkenwald et al., 2022). Combined with the predictions made by Buhmann et al. & Heinrich et al., this project resulted in the largest fully mapped connectome to date with 139 thousand proofread neurons, 2.7 million thresholded connections and a total of 34 million synapses. Among other information, the FlyWire database also includes the atlas of neuropils (brain regions) developed by Ito and colleagues(Ito et al., 2014). The complete dataset is available to download through the *Codex* web-app (Matsliah et al., 2023) (codex.flywire.ai) and *fafbseg-py* Python package (Dorkenwald et al., 2023; Schlegel et al., 2023)(github.com/navis-org/fafbseg-py).

For our study, we utilized the connectivity matrix and the neuron classification table published on the Codex web interface to construct the *neuropil projectome* from the connections between the intrinsic neurons. The neuron skeletons (spatial graph structures) with detected somas were also downloaded from *Codex*, the pre- and postsynaptic sites were attached to it using the *fafbseg* Python package. Additionally, the centroids of the neuropil meshes were downloaded from the *fafbseg* package, which are based on data originally published by Jenett and colleagues (Jenett et al., 2012). This information was needed for constructing the EDR model of the neuropil projectome. We used the latest data release available at the time when beginning our study, snapshot 783 from October 2023.

**Neuropil projectome construction**

We constructed the connectome at the level of neuropils (also referred to as the projectome) using the algorithm developed by the FlyWire project team (Dorkenwald et al., 2023; Lin et al., 2023). This method is based on two assumptions about the information flow between neuropils:

1. The information flow through a single neuron can be expressed probabilistically by taking the fraction of its total synapses present in a given neuropil.
2. Incoming and outgoing information flows through the neuron are considered independent events.

For $N$ neuropils and a single neuron, two separate vectors can be constructed to represent the fractions of incoming and outgoing synapses in the different regions, respectively. The tensorial product of these two vectors yields an $NxN$ matrix, where each element $W_{ij}$ represents the probability of the neuron having incoming information in neuropil *i* and outgoing in neuropil *j*. Summing these matrices for each neuron produces the connectivity matrix of the neuropil projectome, with higher matrix values indicating stronger connections between two neuropils. Given the variability in neuropil sizes and the fact that larger neuropils typically contain more synapses, normalizing the weight matrix appears to be an effective approach for accurately representing overall structural connectivity. In line with previous studies (Ercsey-Ravasz et al., 2013; Horvát et al., 2016), we opted to normalize the weight matrix by its columns. Thus, the normalized element $w_{ij}$ of the weight matrix reflects the probability of information flow between neuropils *i* and *j*.

## Hierarchical clustering method

The *scipy* library offers a vast number of options to perform hierarchical clustering in a bottom-up approach (agglomerating clusters in each step). One example of this family of agglomerative algorithms is Ward's method, which aims to minimize the variance within the clusters and uses the Euclidean distance metric (Murtagh & Legendre, 2014; Ward, 1963).

In our specific case, we have the 75x75 weighted adjacency matrix representing the link lengths ($l_{ij} = -\log w_{ij}$ values) between the nodes in the Drosophila projectome. The data points we chose to cluster are given by 75-dimensional vectors represented by the columns of the matrix. Therefore, in simple terms, the clustering groups together neuropils with similar vectors of incoming information flow.

## Network measures

- Degree of a node in a network gives essentially the number of its neighbors (edges connected to it) (Newman, 2010). Mathematically, it can be expressed using the adjacency matrix $A_{ij}$, which takes the value of 1 if there is a link going from $i$ to $j$, otherwise it is 0.

$$k_i = \sum_{j=1}^{N} A_{ij}$$

In a directed network, this value can be separated based on the direction of links (incoming or outgoing), therefore we can talk about the in-degree and out-degree of a node (Newman, 2010). Similarly, in a weighted network the so-called in- and out-strengths (weighted degrees) can be defined, by using the weighted adjacency matrix in the sum.

- Average (binary/unweighted) path length measures the typical number of links in the shortest paths connecting two nodes in the network (Newman, 2010). In small-world networks, this average path length is relatively short compared to the total number of nodes, indicating that nodes are generally accessible from one another with just a few steps (Watts & Strogatz, 1998).

$$APL = \frac{1}{N(N-1)} \sum_{i \neq j = 1}^{N} d_{ij}$$

- Clustering coefficient of a node can be thought of as the probability that two neighbors of the node are connected (Newman, 2010). Overall in the network, we calculate the average for all nodes.

$$CC = \frac{1}{N} \sum_{i=1}^{N} CC_i = \frac{1}{N} \sum_{i=1}^{N} \frac{1}{k_i(k_i - 1)} \sum_{j,k \in \{i\}} A_{jk}$$

Here *{i}* denotes the neighborhood of node $i$ (subgraph consisting of its neighbors).

- Triangular motifs in a network are subgraphs consisting of 3 nodes. In the directed case, there are a total of 16 possible edge configurations between the 3 nodes. These triangles have been found to be characteristic building blocks of different real world networks (Milo et al., 2002).

- A *k-clique* is a fully connected subgraph consisting of $k$ nodes. In the case of directed networks, fully connected means that there are edges/links between all node pairs in both directions(Newman, 2010).

- Length of a link is defined as: $l_{ij} = -\log w_{ij}$. This approach was already used both in structural (Ercsey-Ravasz et al., 2013; Markov et al., 2013) and functional brain networks (Varga et al., 2024; Wandres et al., 2021). The argument is that the link weight is proportional to the probability of information transfer, so the probability for information to pass on from node *i* to node *j* through node *k* would be: $w_{ij}w_{jk}$. Using the logarithmic form these become additive: $-\log w_{ik} = -\log w_{ij} - \log w_{jk} = l_{ij} + l_{jk}$, and we can calculate the shortest paths (weighted distances) in the network that will indicate the most probable paths for information transfer (Ercsey-Ravasz et al., 2013; Markov et al., 2013).

- Distance (or resistance, $r_{ij}$) between two nodes in the network is the length of the shortest path. Its inverse is sometimes called "conductance" (in analogy with physical circuits)(Ercsey-Ravasz et al., 2013; Markov et al., 2013).

- Global communication efficiency is defined as the average of the inverse resistance ("conductance"), here defined as length of shortest path between all node pairs (Latora & Marchiori, 2003):

$$E_g = \frac{1}{N(N-1)} \sum_{i \neq j=1}^{N} \frac{1}{r_{ij}}$$

- Local communication efficiency is (Vragović et al., 2005):

$$E_l = \frac{1}{N} \sum_{i=1}^{N} E_{l_i} = \frac{1}{N} \sum_{i=1}^{N} \frac{1}{k_i(k_i - 1)} \sum_{j,k \in \{i\}} \frac{1}{r_{jk/i}}$$

where *{i}* indicates the set of neighbors of node *I*; *j* and *k* are neighbors of *I*; $r_{jk/i}$ is the shortest path between *j* and *k* after the removal of node *i* and its links from the graph; $k_i$ is the degree (number of neighbors, links) of node *i*. This gives information on how efficient is the communication between the neighbors of node *i*, but without using the links connecting them to node *i*. This local measure is averaged over all nodes (the first sum in the formula).

- Backbone of a weighted network is obtained by sequentially removing its weakest links while maintaining network connectivity (Ercsey-Ravasz et al., 2013). This means that there remains at least one path between every pair of nodes. For directed networks, this connectivity can be defined in two ways: a *strongly connected backbone*, where paths must exist in both directions between all pairs of nodes, or a *weakly connected backbone*, where the direction of edges is ignored.

- The Kamada-Kawai layout is a method for visualizing networks in space using a force-directed algorithm. In this approach, the network is modeled as a system where each pair of nodes is connected by a virtual spring with a strength inversely proportional to the square of the graph-theoretic distance between them(Kamada & Kawai, 1989). This algorithm seeks to minimize the total energy of the system by adjusting the positions of the nodes so that the Euclidean distances between them closely match the desired graph distances. As a result, nodes that are close within the network (i.e., connected by short paths) are placed near each other in the visualization, while nodes that are distant are positioned further apart.

- The weight asymmetry of a link between two nodes $i$ and $j$ is defined as the relative difference between the weights of the links in both directions:

$$ASYM_{ij} = \frac{|w_{ij} - w_{ji}|}{w_{ij} + w_{ji}}$$

This metric quantifies the degree of asymmetry in the weights of bidirectional links between nodes. It ranges from 0 to 1, a lower value indicates stronger symmetry, and a higher one greater asymmetry. It can be also defined for unidirectional links, in which case this measure is equal to 1.

- Based on the previous asymmetry metric, a new network can be constructed describing asymmetries with unidirectional links only: if $w_{ij} > w_{ji}$ we insert a directed link pointing from $i$ to $j$ with a weight of $ASYM_{ij}$; in the other case the direction of the link is the opposite (from $j$ to $i$). This approach ensures that the direction of the link corresponds to the predominant direction of connection between the nodes. Then, the overall strength of asymmetries for a node can be measured by computing the in- and out-strengths (weighted degrees) in this newly formed network. Nodes that have mainly out-going links (large out-strength of asymmetries) are at the bottom of functional hierarchy (e.g. optic lobe): they obtain information from sensory inputs outside the brain and forward it to more higher-level (central) areas for processing, those have mainly incoming connections.

**Other statistical measures**

- Root mean square deviation (RMSD) of two different samples is defined as:

$$RMSD = \sqrt{\frac{1}{n}\sum_{i=1}^{n}(x_i - y_i)^2}$$

In this context, RMSD is used to measure the differences between samples (properties of the brain network) derived from the real dataset and those predicted by the model.

**Software packages**

The database of the Drosophila connectome was downloaded from the Codex web interface (Matsliah et al., 2023) and fafbseg python package (Dorkenwald et al., 2023; Schlegel et al., 2023). A large part of the network analysis was performed using igraph library (Csárdi et al., 2024), for visualization the NetworkX package was used (Hagberg et al., 2008). Neuron skeletons were processed with navis package (Philipp Schlegel et al., 2024), the neuropil plots were created in the neuroglancer environment (Maitin-Shepard et al., 2021). Hierarchical clustering was performed using the scipy implementation (Virtanen et al., 2020) and our own code for EDR model network generation is available on GitHub (https://github.com/bpentek/EDRmodel).

## Supporting Information

**The exponential distance rule based network model predicts topology and reveals functionally relevant properties of the Drosophila projectome**

Balázs Péntek, Mária Ercsey-Ravasz

## Supplementary Figures

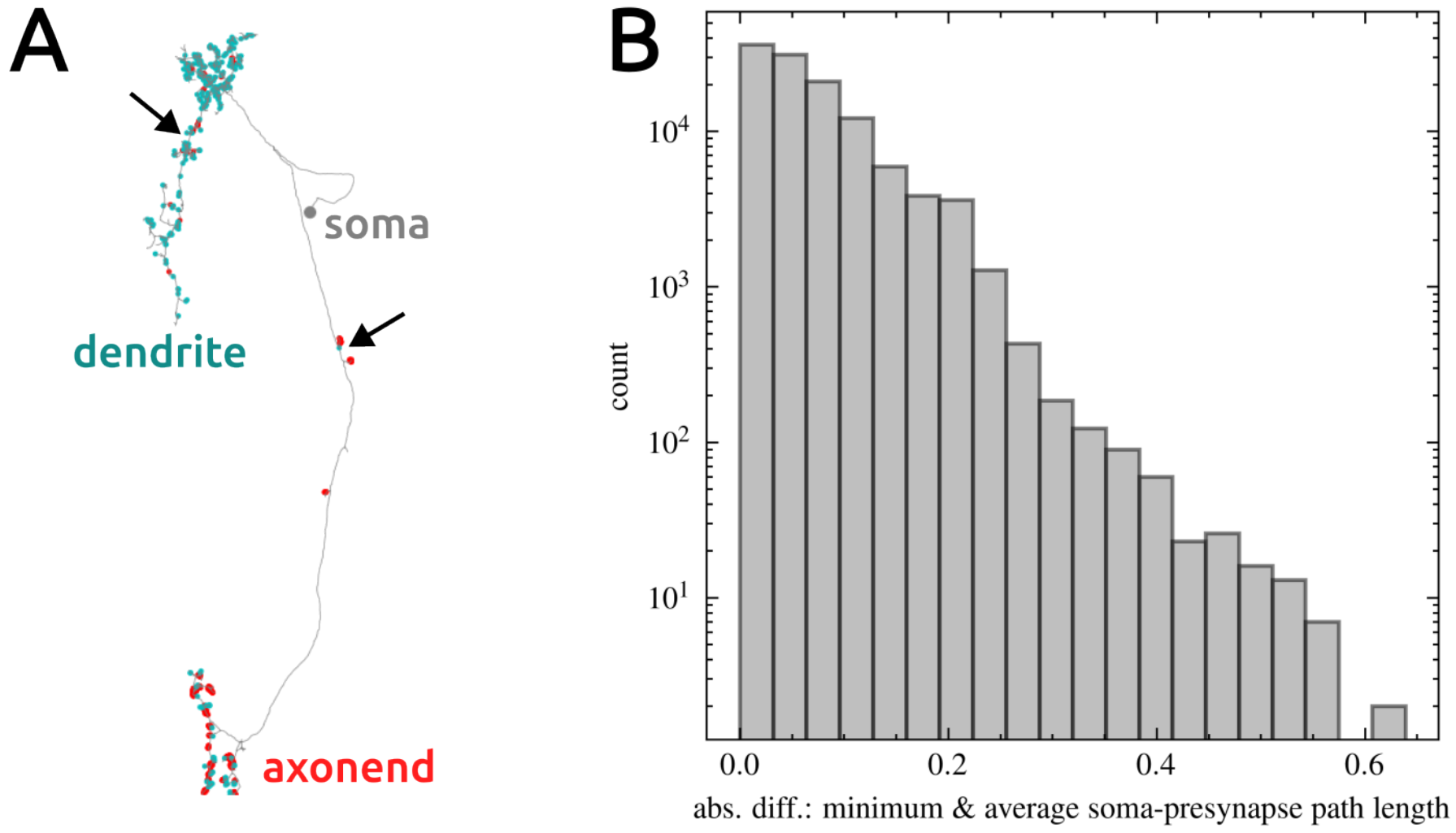

**Supplementary Figure 1. Excluding errors and special synapses.** A) An example for non-traditional presynaptic points shown on the axon/dendrite (black arrows pointing to the red dots). B) Distribution of the absolute difference between minimum soma-presynapse path length and the average.

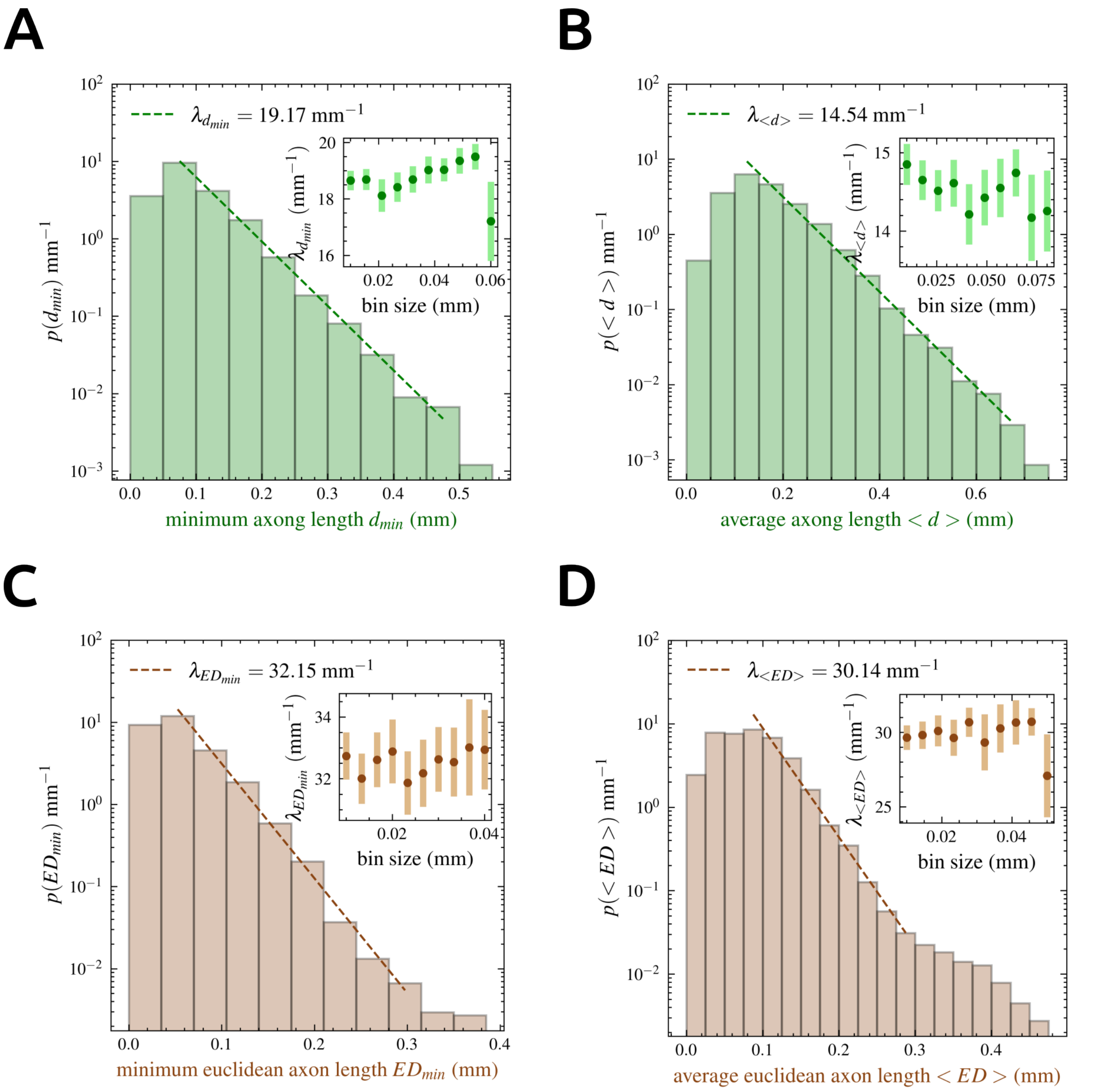

**Supplementary Figure 2. The EDR distributions for all neurons without using a threshold for the absolute difference between minimum and average soma-presynapse path length.** A) Probability distribution of minimum path length, B) average path length, C) minimum Euclidean distance, D) average Euclidean distance. Using different bin sizes we estimate the following intervals for $\lambda$: A) [15.8, 19.95] mm$^{-1}$ B) [13.6, 15.1] mm$^{-1}$ C) [30.8, 34.6] mm$^{-1}$ D) [24.3, 32.2] mm$^{-1}$.

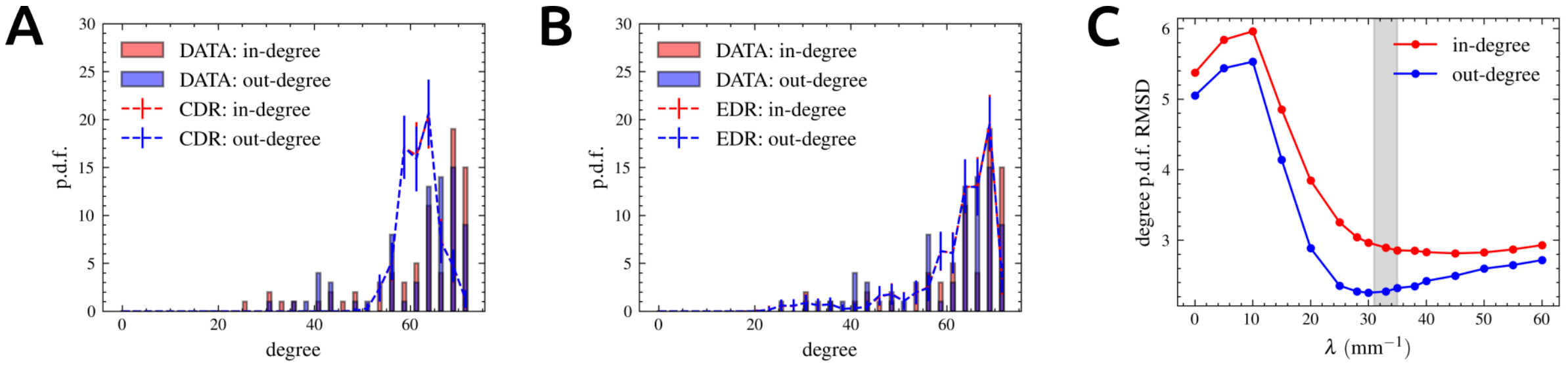

**Supplementary Figure 3. Degree distributions: dataset vs. model.** Having a small network with 75 nodes, these are relatively noisy. A) in- and out-degree (red and blue) of dataset (bar plots) and average values for CDR models with $\lambda = 0\ mm^{-1}$(dashed line; errorbar indicating standard deviation); B) similar to A), but for EDR model with $\lambda = 33\ mm^{-1}$; C) RMSD as a function of $\lambda$, the gray interval showing the fitted values in Fig. 1F: $\lambda$=[31, 35]$mm^{-1}$.

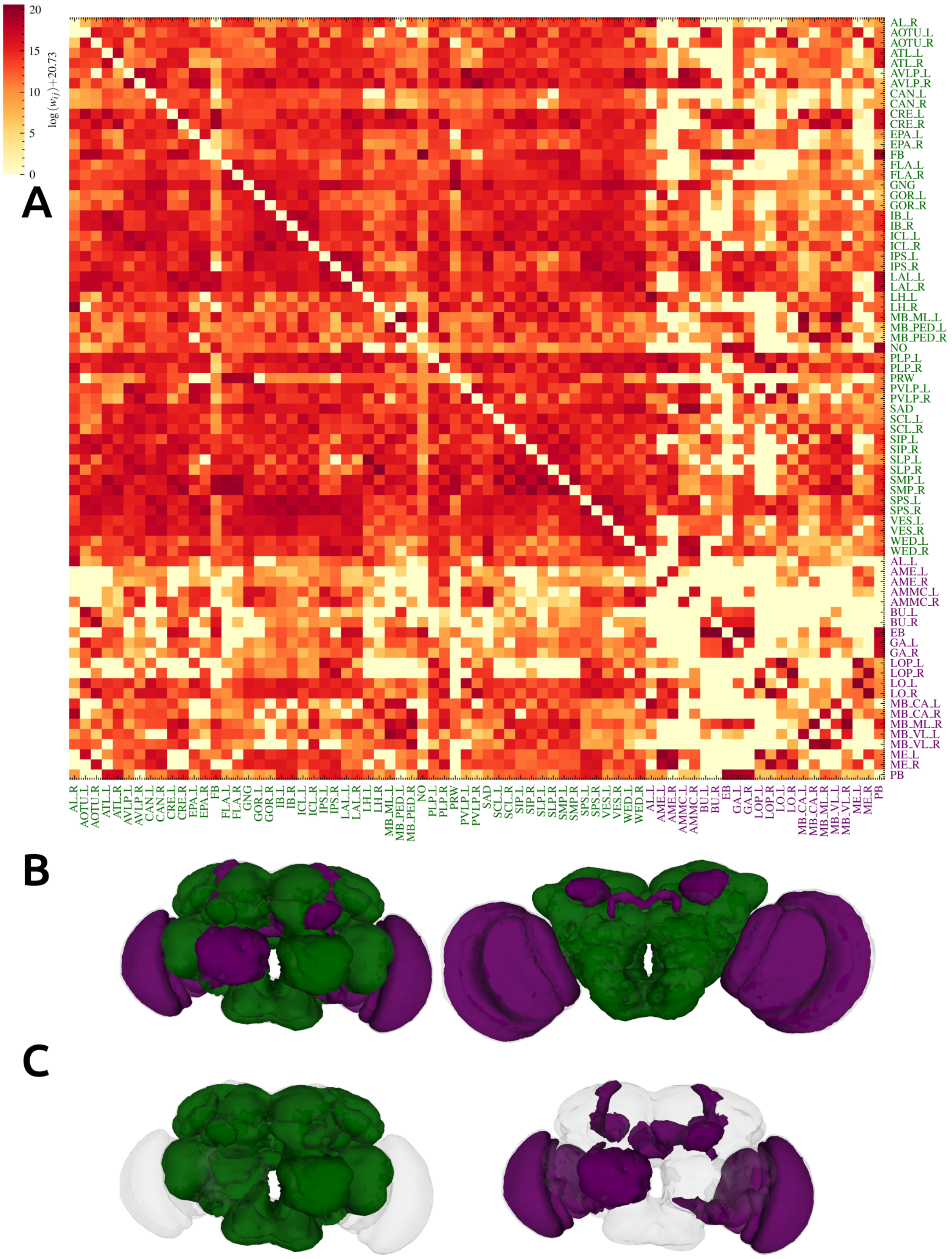

**Supplementary Figure 4. Core-periphery structure based on the largest cliques in the Drosophila projectome.** A) Matrix plot representing connection strength between neuropils. The rows and columns are ordered such as the nodes corresponding to the network core are shown first (green labels), then the periphery (purple labels). The network core consists of 53 nodes in total, resulting from the set of nodes in the 31 largest cliques of size 43. We can see the large density and strong connections inside the core. B) The core (green) and peripheral (purple) neuropils shown on the Drosophila brain from a front and back view. C) The same two sets of neuropils shown separately on the Drosophila brain from a front view (core on the left, periphery on the right).

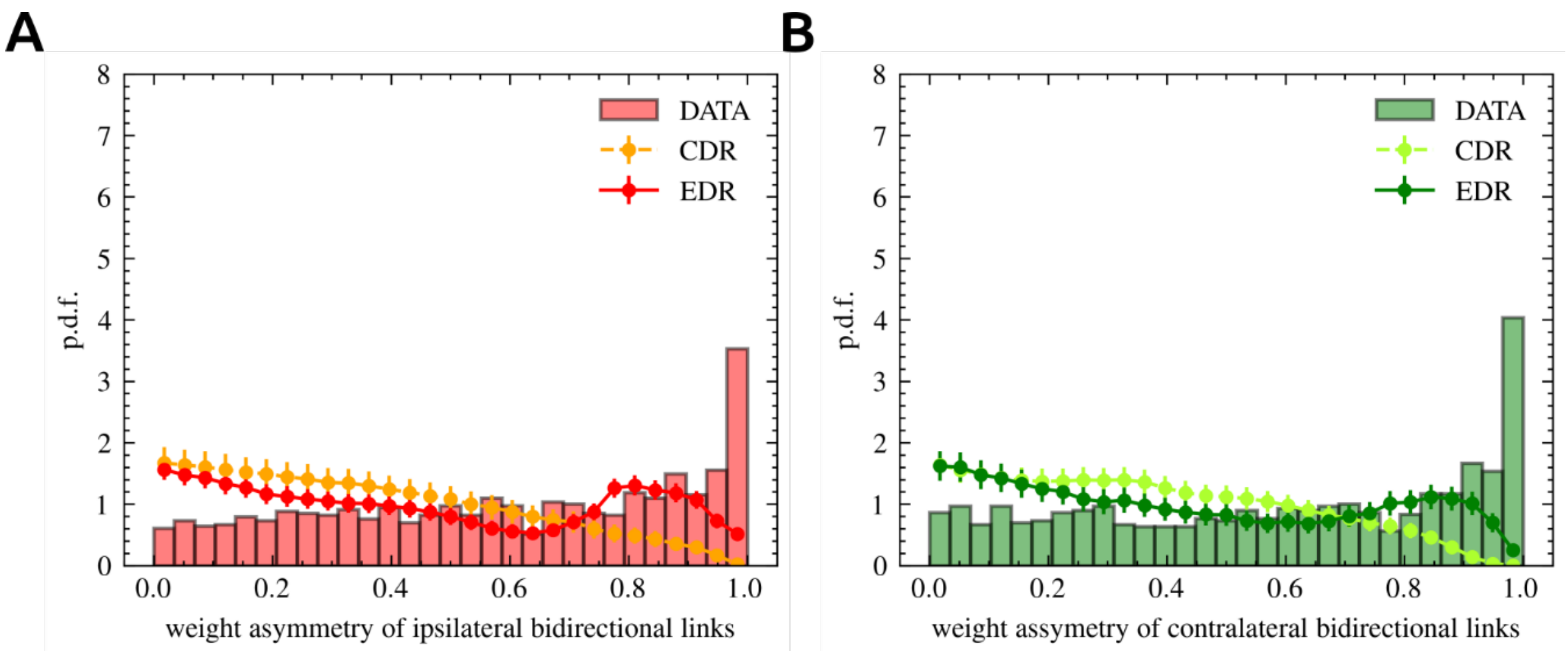

**Supplementary Figure 5. Bidirectional weight asymmetry distribution.** A) Weight asymmetry distribution for ipsilateral bidirectional links, and B) contralateral bidirectional links.

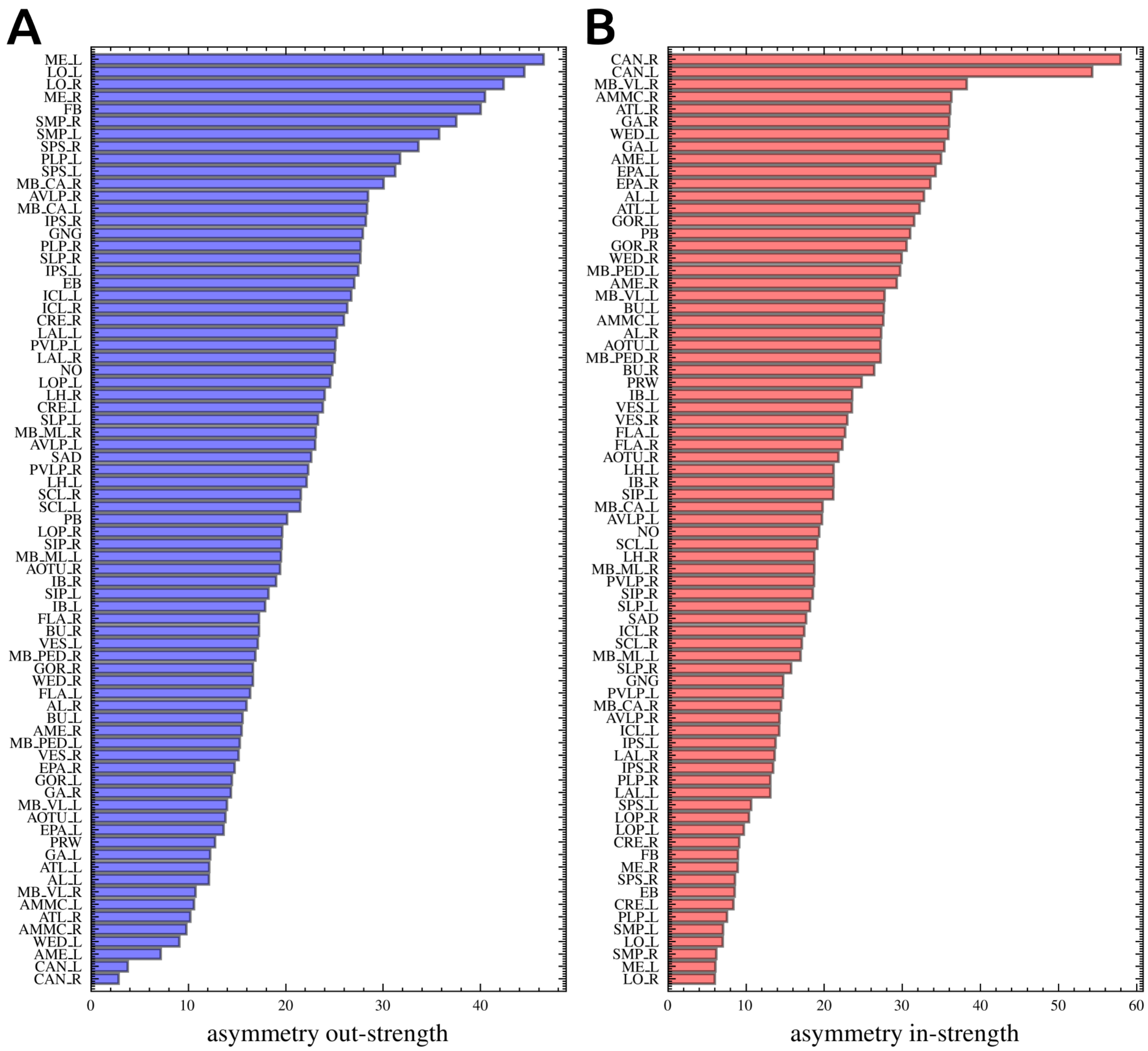

**Supplementary Figure 6. Order of areas based on link asymmetry strengths.** A) Brain regions are ordered based on the out-strength and B) in-strength of asymmetries.

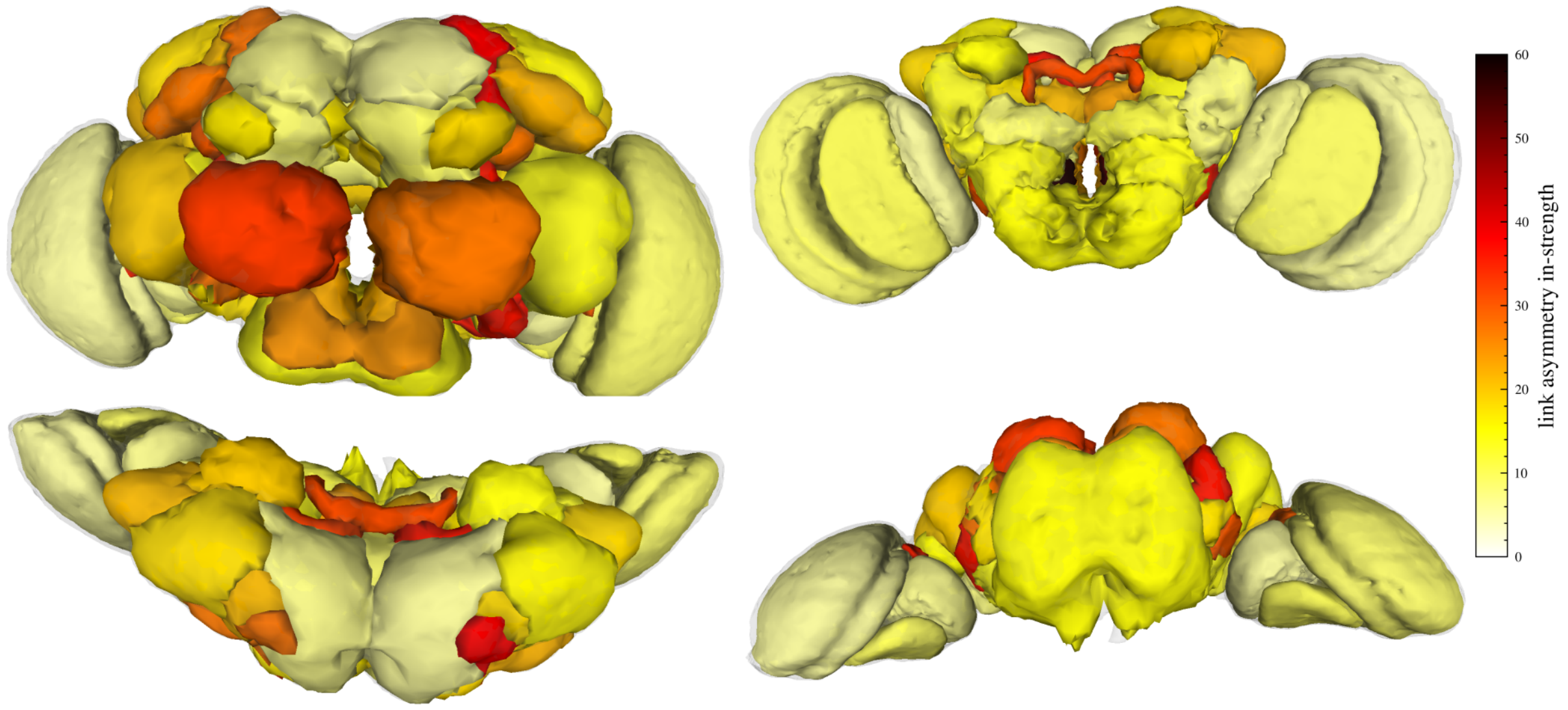

**Supplementary Figure 7**. **Brain map colored based on asymmetry in-strengths** (values from Supplementary Figure 6B). Front, back, top and bottom view.